\documentclass[showpacs,prl,twocolumn,floatfix,reprint]{revtex4}
\usepackage{graphicx,amssymb,amsmath,xcolor}
\usepackage{stmaryrd}

\begin{document}

\title{Central loops in random planar graphs} 
%The spatial patterns of betweenness centrality}

\author{Benjamin Lion}
\affiliation{Institut de Physique Th\'{e}orique, CEA, CNRS-URA 2306, F-91191, 
Gif-sur-Yvette, France}

\author{Marc Barthelemy}
\email{marc.barthelemy@cea.fr}
\affiliation{Institut de Physique Th\'{e}orique, CEA, CNRS-URA 2306, F-91191, 
Gif-sur-Yvette, France}
\altaffiliation{CAMS (CNRS/EHESS) 190-198, avenue de France, 75244 Paris Cedex 13, France}

\begin{abstract}

  Random planar graphs appear in a variety of context and it is
  important for many different applications to be able to characterize
  their structure. Local quantities fail to give interesting
  information and it seems that path-related measures are able to
  convey relevant information about the organization of these
  structures. In particular, nodes with a large betweenness centrality
  (BC) display non-trivial patterns, such as central loops. We first
  discuss empirical results for different random planar graphs and we
  then propose a toy model which allows us to discuss the condition
  for the emergence of non-trivial patterns such as central
  loops. This toy model is made of a star network with $N_b$ branches
  of size $n$ and links of weight $1$, superimposed to a loop at
  distance $\ell$ from the center and with links of weight $w$. We
  estimate for this model the BC at the center and on the loop and we
  show that the loop can be more central than the origin if $w<w_c$
  where the threshold of this transition scales as $w_c\sim n/N_b$. In
  this regime, there is an optimal position of the loop that scales as
  $\ell_{opt}\sim N_b w/4$. This simple model sheds some light on the
  organization of these random structures and allows us to discuss the
  effect of randomness on the centrality of loops. In particular, it suggests that
  the number and the spatial extension of radial branches are the
  crucial ingredients that control the existence of central loops.

\end{abstract}


\maketitle

\section{Introduction}

Random planar graphs -- random graphs that can be drawn on the 2d
plane with no edge crossing \cite{Clark:1991} -- pervade many
different fields from abstract mathematics
\cite{Tutte:1963,Bouttier:2004}, to theoretical physics
\cite{Ambjorn:1997}, botanics \cite{Weitz:2012,Katifori:2012},
geography and urban studies \cite{Barthelemy:2011}. In particular,
planar graphs are central in biology where they can be used to
describe veination patterns of leaves or insect wings and which
display an interesting architecture with many loops at different
scales \cite{Weitz:2012,Katifori:2012,Katifori:2010,Hu:2013}. In the
study of urban systems, planar networks are extensively used to
represent, to a good approximation, various infrastructure networks
\cite{Barthelemy:2011} such as transportation networks
\cite{Haggett:1969} and streets patterns
\cite{Hillier:1984,Marshall:2006,Jiang:2004,Roswall:2005,Porta:2006,Porta:2006b,Lammer:2006,Crucitti:2006,
  Cardillo:2006,Xie:2007,Jiang:2007,Masucci:2009,Chan:2011,Courtat:2011,Strano:2012,Barthelemy:2013,
  Viana:2013,Louf:2014,Porta:2014}. Understanding the structure and
the evolution of these networks is therefore interesting from a purely
graph theoretical point of view, but could also have an impact in
different fields where these structures are central.

Most previous studies characterize different aspects of these graphs,
either purely topological (degree distribution, clustering, etc.) or
geometrical (angles, segment length, face area distribution,
etc.). Due to spatial constraints, most local information such as the
degree distribution, the clustering or the assortativity have however
a trivial behavior \cite{Barthelemy:2011}. In addition the important
information about these random planar graphs is in fact not in their
adjacency matrix only but also in their geometry described by the
spatial distribution of nodes and relevant meaures should combine
topology and geometry.  Despite the large number of studies on these
graphs, there is still a lack of a non-local high-level metrics that
allow for understanding and comparing these graphs with each
other. However, a promising direction is given by path-based
quantities such as the simplicity \cite{Aldous:2010,Viana:2013} or as
we will discuss here, the betweenness centrality (BC)
\cite{Freeman:1977}. The BC was introduced to quantify the importance
of a node (or an edge) in a network, but it also proved to be a very
interesting tool in the study of random planar graphs. Already in
\cite{Lammer:2006}, interesting spatial patterns of nodes with large
BC were observed. More recently, it has been shown
that the most salient aspects of the structural changes during the
evolution of the street network of Paris \cite{Barthelemy:2013} is
revealed by the spatial distribution of the nodes with the largest
BC. These different results therefore point to the fact that high
centrality nodes form non-trivial patterns, and among them, the
emergence of loops. All these different results point to the fact the
BC, which is relatively simple, is a good candidate for monitoring and
understanding the organization of random planar graphs. In this study
we will focus on the apearance of loops made of links with large BC
and we will propose a simple toy model that allows us to discuss the
conditions for the appearance of such patterns.

The emergence of rings in largely urbanized areas is a common fact and
the study presented here gives a topological light on this
phenomenon. Our study echoes previous work where congestion effect at
a central hub could be so high that avoiding the ring is beneficial
\cite{Ashton:2005,Jarrett:2006}. Here, in contrast, we do not take
into account congestion and discuss the conditions necessary for a
loop to become more interesting in terms of time cost.

\section{The BC for planar graphs}

Basic results on planar networks can be found in any graph theory
textbook (see for example \cite{Clark:1991} and for useful algorithms
see \cite{Jungnickel:2008}) and we will very briefly recall the
definition of these objects.  Basically, a planar graph is a graph that can be
drawn in the plane in such a way that its edges do not intersect.  Not
all drawings of planar graphs are without intersection and a drawing
without intersection is sometimes called a plane graph or a planar
embedding of the graph. In real-world cases, these considerations
actually do not apply since the nodes and the edges represent in
general physical objects. More precisely, we will focus here on the case of planar
graphs embedded in 2d space, which typically describes systems such as
the road and street network. 

\subsection{Definition of the BC and variants}

The betweenness centrality counts the fraction of shortest paths going
through a given node (or link) and is given by \cite{Freeman:1977}
\begin{equation}
g(v)={\cal N}\sum_{s\neq t}\frac{\sigma_{st}(v)}{\sigma_{st}}
\label{eq:bc}
\end{equation}
where $v$ is a node, $\sigma_{st}$ is the number of shortest paths from $s$ to $t$
and $\sigma_{st}(v)$ those paths going through $v$. In general the
summation is on $s\neq t$ and $s\neq v$, $t\neq v$ and this is the
convention that we will adopt in this paper. The constant ${\cal N}$ is the
normalisation and we will use here ${\cal N}=1/(N-1)(N-2)$ which counts the
number of pairs and ensures that $g(v)\in [0,1]$. For edges, the
definition of the BC is similar to Eq.~\ref{eq:bc}.

It is important to stress here that the BC could in fact be defined
for any type of paths. The most common choice is the shortest path
but we will use the more general case of weighted shortest path, which
corresponds to the quickest path if the weight of a link represents
time. For numerical calculations, we implemented the now standard algorithm of
Brandes \cite{Brandes:2001}.

\subsection{Regular lattice}

\begin{figure}[!h]
\centering
\includegraphics[scale=0.25]{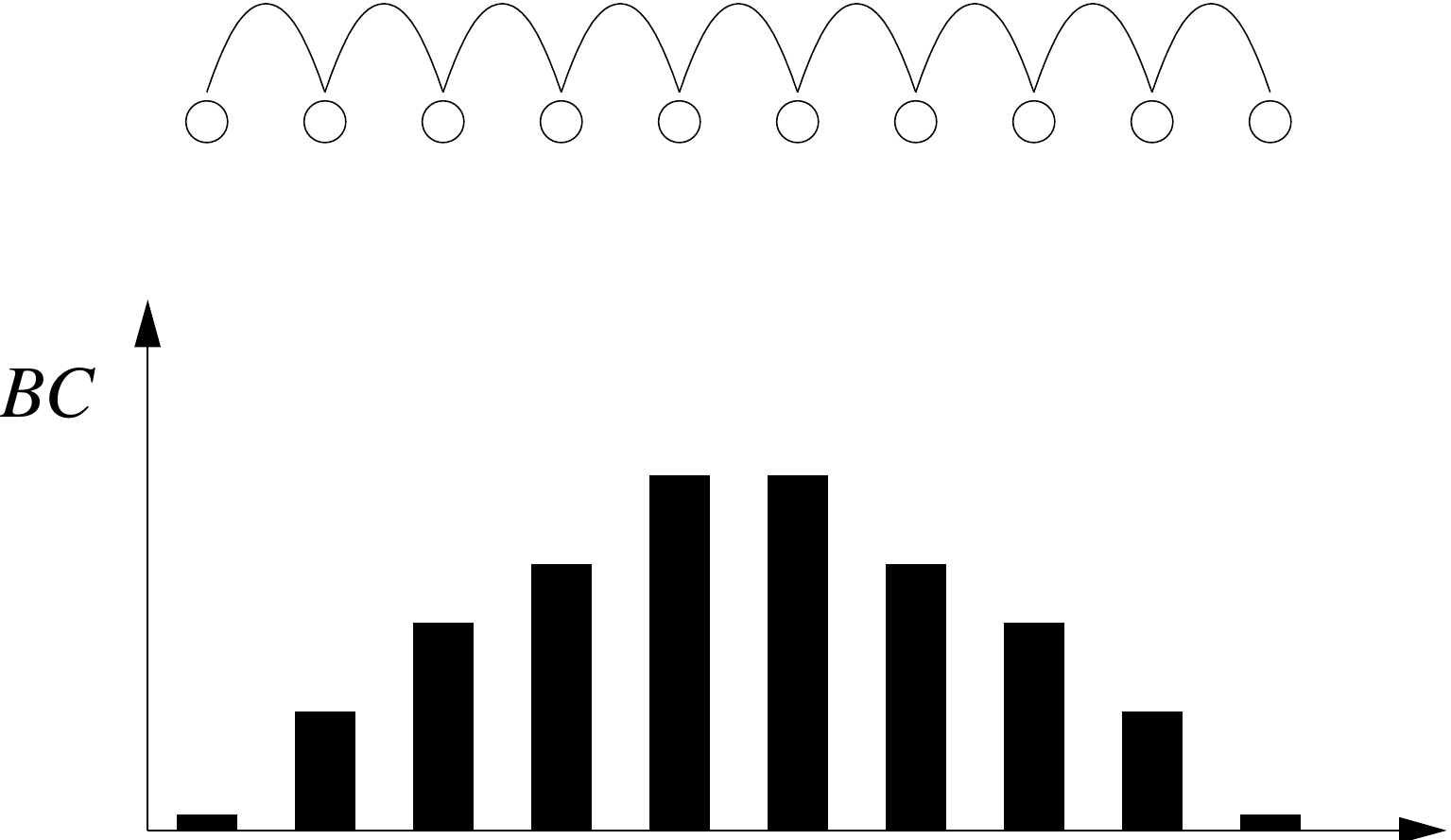}
\includegraphics[scale=0.25]{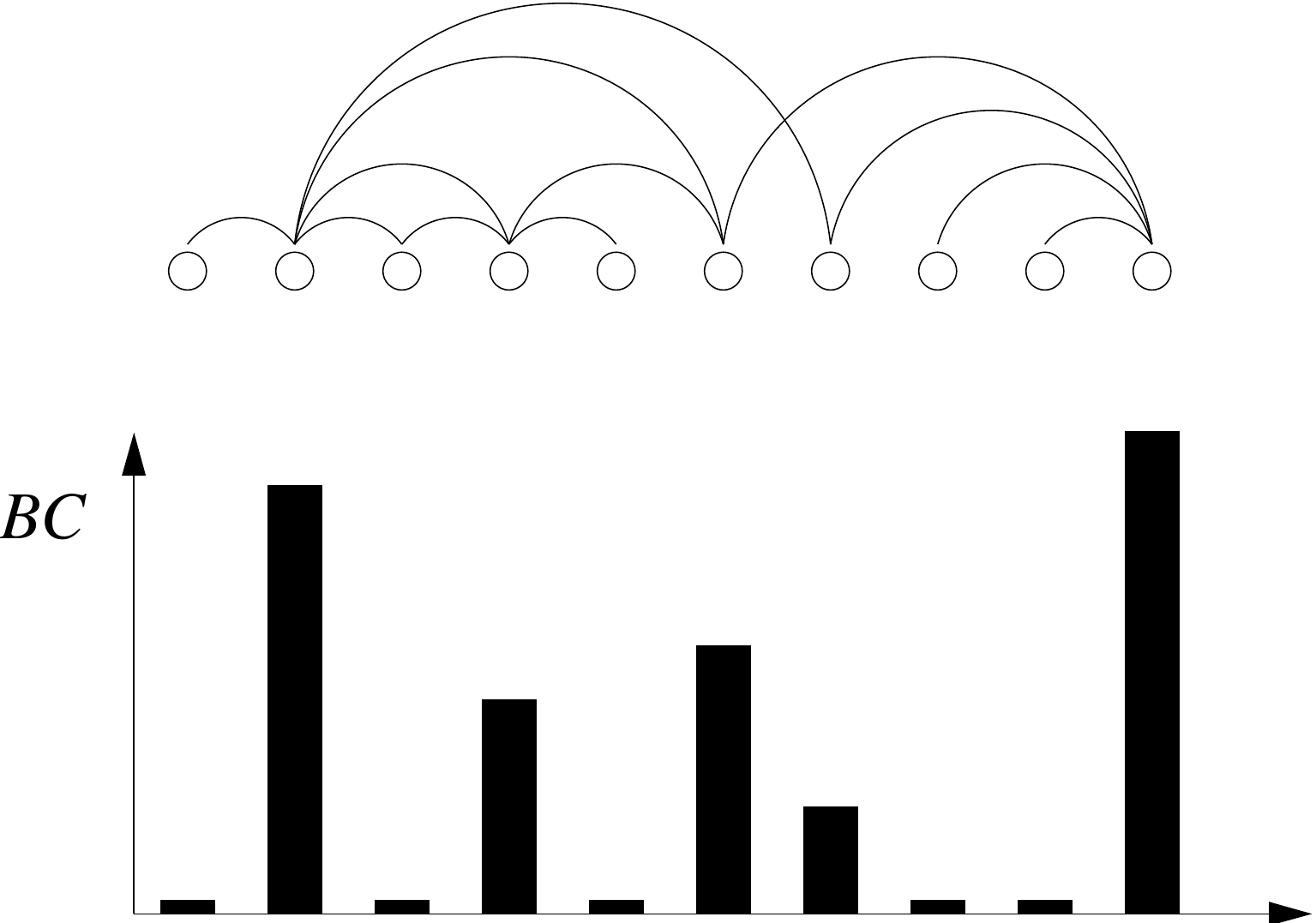}
\caption{ Betweenness centrality for a one-dimensional lattice. (Left)
  When there is no disorder, the barycenter is the most central
  nodes. (Right) In the case of a disordered network, degree becomes
  relevant and the most central nodes have large degrees.
\label{fig:nat_vs_lat}}
\end{figure}

In a one-dimensional lattice of size $n$, the BC of a site $0\leq x\leq
n$ is given by
\begin{equation}
g(x)=x(n-x)
\end{equation}
and the maximum thus corresponds to the barycenter of all nodes (see
Fig.\ref{fig:nat_vs_lat}) (in the two-dimensional case we obtain a
similar behavior).
When we introduce disorder - by removing or rewiring links - the BC
becomes important at nodes that can be far away from the barycenter (see
Fig.\ref{fig:nat_vs_lat}). In the extreme case where space doesn't
play a role anymore such as in scale-free networks, the average BC per degree classes $g(k)$
scales as \cite{Barthelemy:2004}
\begin{equation}
g(k)\sim k^\eta
\end{equation}
where $\eta$ is an exponent that depends on the structure of the
graph. Even if there are fluctuations around this scaling it shows
here that essentially the degree controls the BC in these graphs.

\subsection{Percolation: giant component}

\begin{figure}[!h]
\includegraphics[scale=0.4]{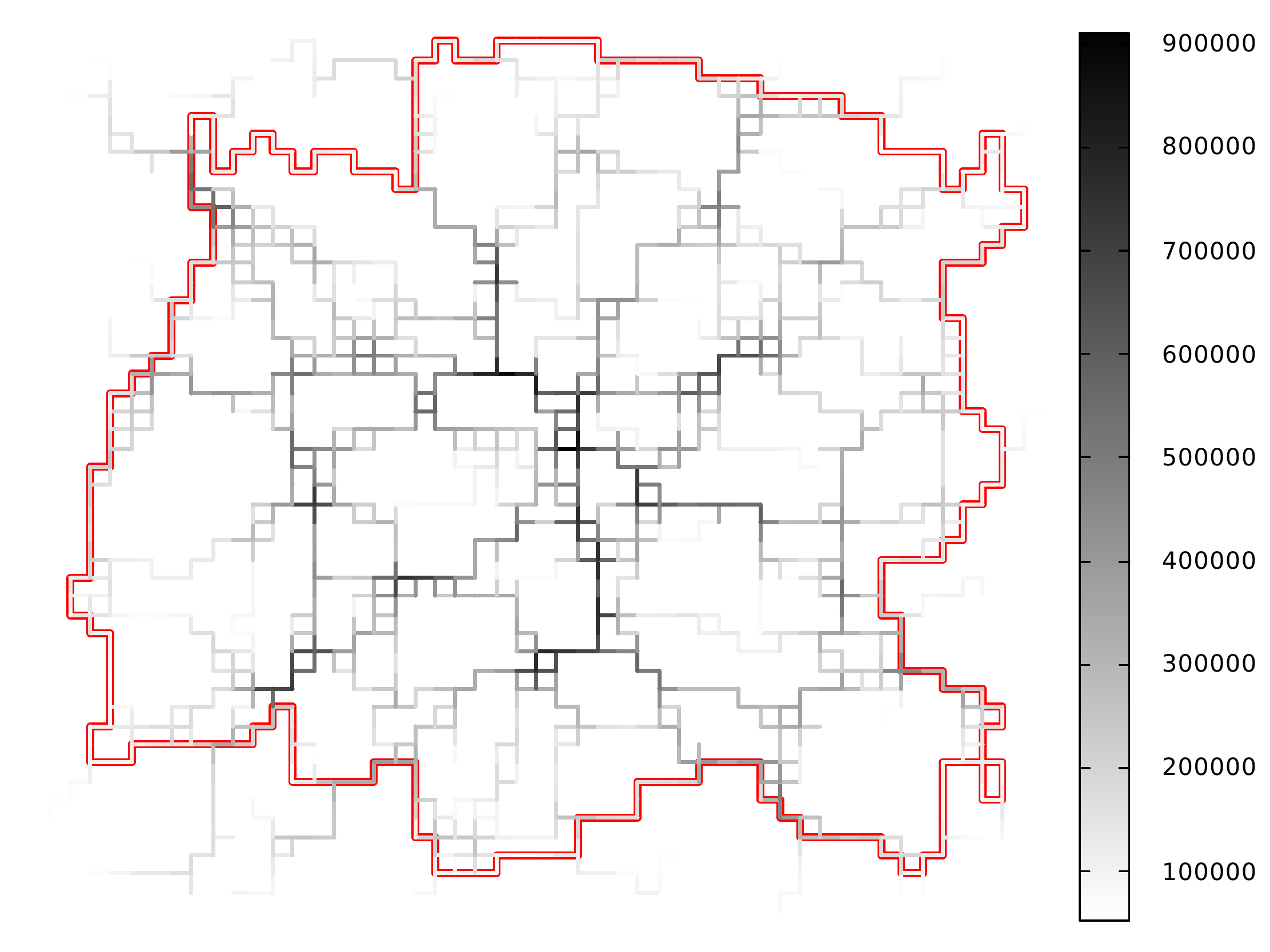}
\caption{ Links belonging to the giant component obtained for
  percolation with $f=0.15$ and with normalized BC larger than
  $g^*=0.05g_{max}$. We observe that the links with a very large BC
  form a non-trivial pattern and are not necessarily close to the
  center. The largest loop is here highlighted with a double line.}
\label{fig:perco}
\end{figure}

A simple way to construct a random planar graph is to consider a
regular lattice where each link has a probabilty $f$ to be removed
(and $p=1-f$ to be present). Above the percolation threshold
($p\geq p_c$), the system displays a giant component which connects a
non-zero fraction of the nodes. We can study the BC on this giant
component and filter them for different threshold $g^*$: we keep
only links with centrality $g$ such that $g>g^*$.  We show
in Fig.~\ref{fig:perco} the set of links that belong to the giant
component and with BC larger than $g^*$ and represent the BC with a
color code (from dark blue to yellow).
and we observe that the set of most central links forms a
non-trivial pattern where the distance to the center is not the main
determinant (Fig.~\ref{fig:perco}). In particular, we observe the
presence of very central links that are not close to the center and
that depend on the particular disorder configuration.

\begin{figure}[!h]
\includegraphics[scale=0.3]{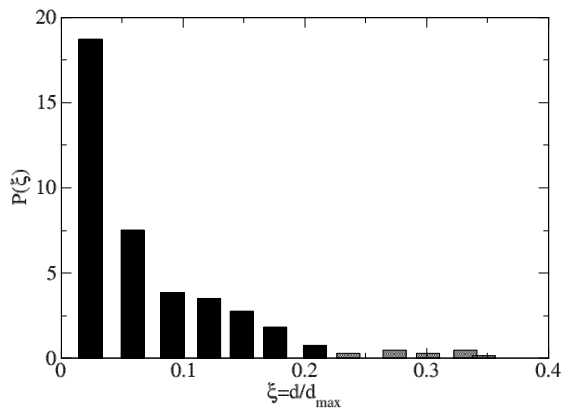}
\caption{ Distribution of the distance (normalized by the maximum
  distance on the lattice) to the center for nodes with a ratio
  $\eta>1$. These results were obtained for $p=0.8$ on a $100\times
  100$ lattice and averaged over $30$ configurations.}
\label{fig:pd}
\end{figure}

We can go further in the analysis of the structure of the percolating
cluster by analyzing the ratio 
\begin{equation}
\eta=\frac{g(r,\theta)}{\max_{r'<r,\theta'\in [0,2\pi]}g(r,\theta')}
\end{equation}
which compares the BC at one point with the maximum BC of nodes
in the region closer to the origin. For a percolating
cluster obtained at $p=0.8$ (well above the percolation threshold) and
on a lattice $100\times 100$, we 
observe a very broad distribution of $\eta$. For values larger than
one we obtain on average $\overline{\eta}\simeq 3$ and a very large dispersion of order
$10^3$. We can observe the points for which we have a ratio $\eta>1$
and plot (Fig.~\ref{fig:pd}) the distribution of the distance to the center for these
points (normalized by the maximum distance $d_{max}$).
This figure \ref{fig:pd} shows that the location of nodes with a very large BC (at
least larger than the BC of the nodes closer to the center) can be of
order the system size. This shows that -- depending on the disorder --
the `central' area composed of the geometrical center and its
surroundings are composed of nodes with a relatively small BC. This
reinforces the need to understand in which cases the monotoneous
decrease of the BC with the distance to center can be strongly
modified by fluctuations.

\subsection{Real-world planar graphs}

Streets and roads form a network where nodes are intersections and
links are segment roads, and which is planar (or almost planar, to a
good approximation). This network is now fairly well characterized and
due to spatial constraints, the degree distribution is peaked, the
clustering coefficient and assortativity are large, and most of the
interesting information lies in the spatial distribution of
betweeenness centrality \cite{Barthelemy:2011}. Many studies
\cite{Jiang:2004,Roswall:2005,Porta:2006,Porta:2006b,Lammer:2006,Crucitti:2006,Cardillo:2006,Xie:2007,Jiang:2007,
  Masucci:2009,Chan:2011,Courtat:2011,Strano:2012,Barthelemy:2013,Porta:2014}
considered different aspects of this network and observed non-trivial
structures in the BC spatial distribution. In particular, in
\cite{Lammer:2006} it has been observed that the distribution of the BC can
display non-trivial spatial patterns and in \cite{Barthelemy:2013} the
authors showed that during the evolution of the street network of
Paris (France) most `standard' measures were unable to detect the
important structural changes that occurred in the $19^{th}$ century, while in
contrast, the spatial distribution of the BC displayed dramatic changes.

\begin{figure}[!h]
\centering
\includegraphics[scale=0.4]{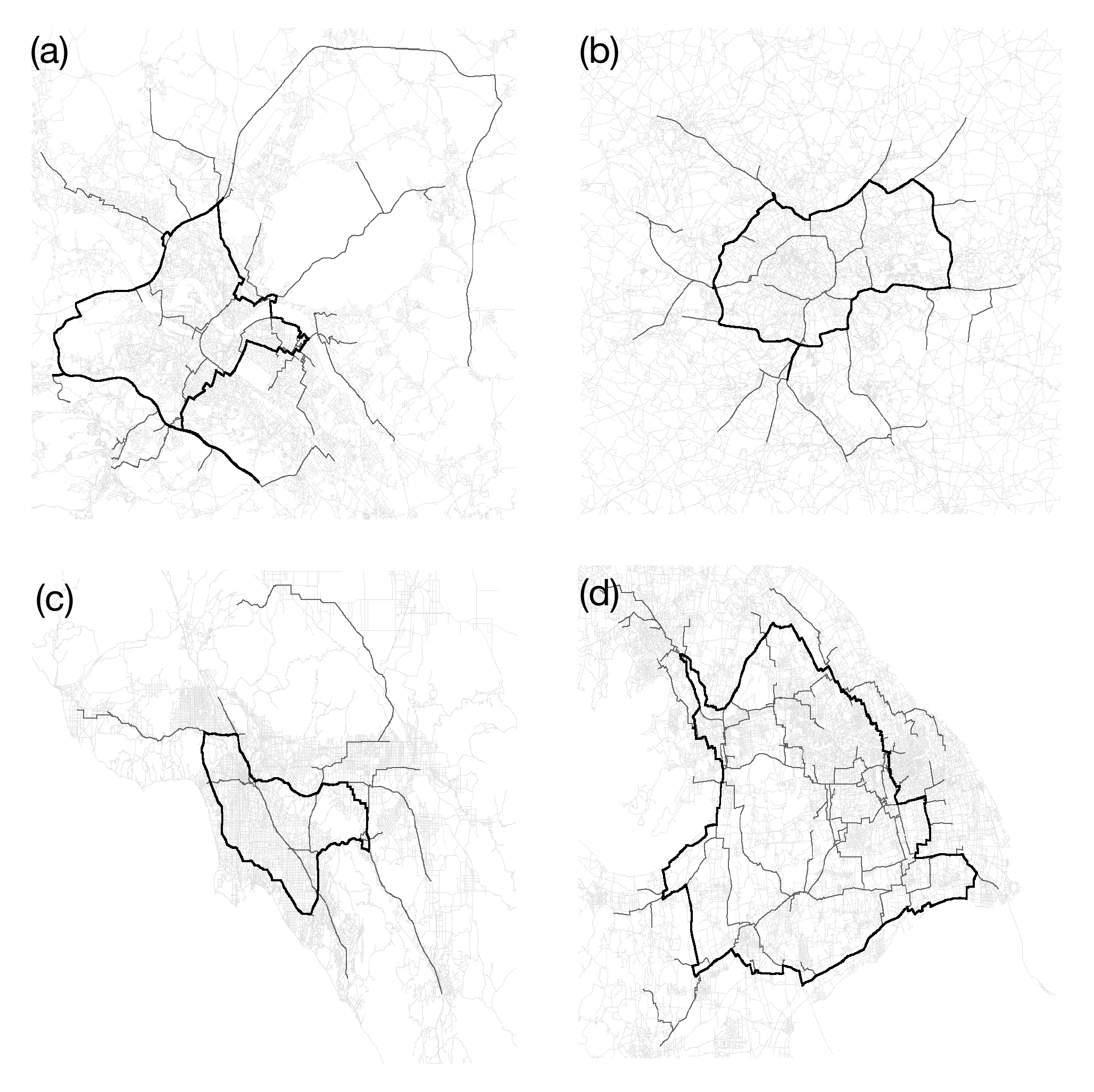}
\caption{ We show for real-world networks the links $e$ with a BC larger than a certain
  threshold $g(e)>g^*$ and we highlight the largest loop. (a) Dresden,
  Germany ($g^*=0.11$). (b) Paris, France ($g^*=0.315$). (c) Los Angeles, USA ($g^*=0.05$). (d) Shanghai ($g^*=
  0.07$).
  \label{fig:realworld}}
\end{figure}

\begin{figure}[!h]
\centering
\includegraphics[scale=0.30]{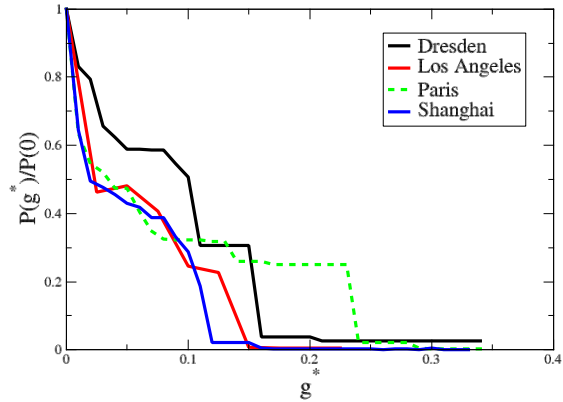}
\caption{ Perimeter $P(g^*)$ of the main loop (normalized by perimeter of the
  loop at $g^*=0$) for the road network of Dresden, Los Angeles, and
  Paris.
\label{fig:perimeter}}
\end{figure}

Using the road network obtained from city extracts (the data has been obtained from the
Mapzen website \cite{Mapzen}), we compute the BC distribution for
different cities shown in Fig. \ref{fig:realworld}.
For all these real-world cases, we observe that indeed non-trivial
structures appear and in particular we observe the appearance of loops
made of central links and of different sizes. We can
test the stability of these loops, by filtering these networks for
different values of the BC threshold $g^*$ and compute the perimeter
of the main loop. The results for Dresden, Los Angeles, and Paris are shown in Fig.~\ref{fig:perimeter}.
We observe on this plot the presence of various plateaus at
intermediate values of $g^*$ suggesting
that these loops are indeed very central and stable. 

\begin{figure}[!h]
\centering
\includegraphics[scale=0.35]{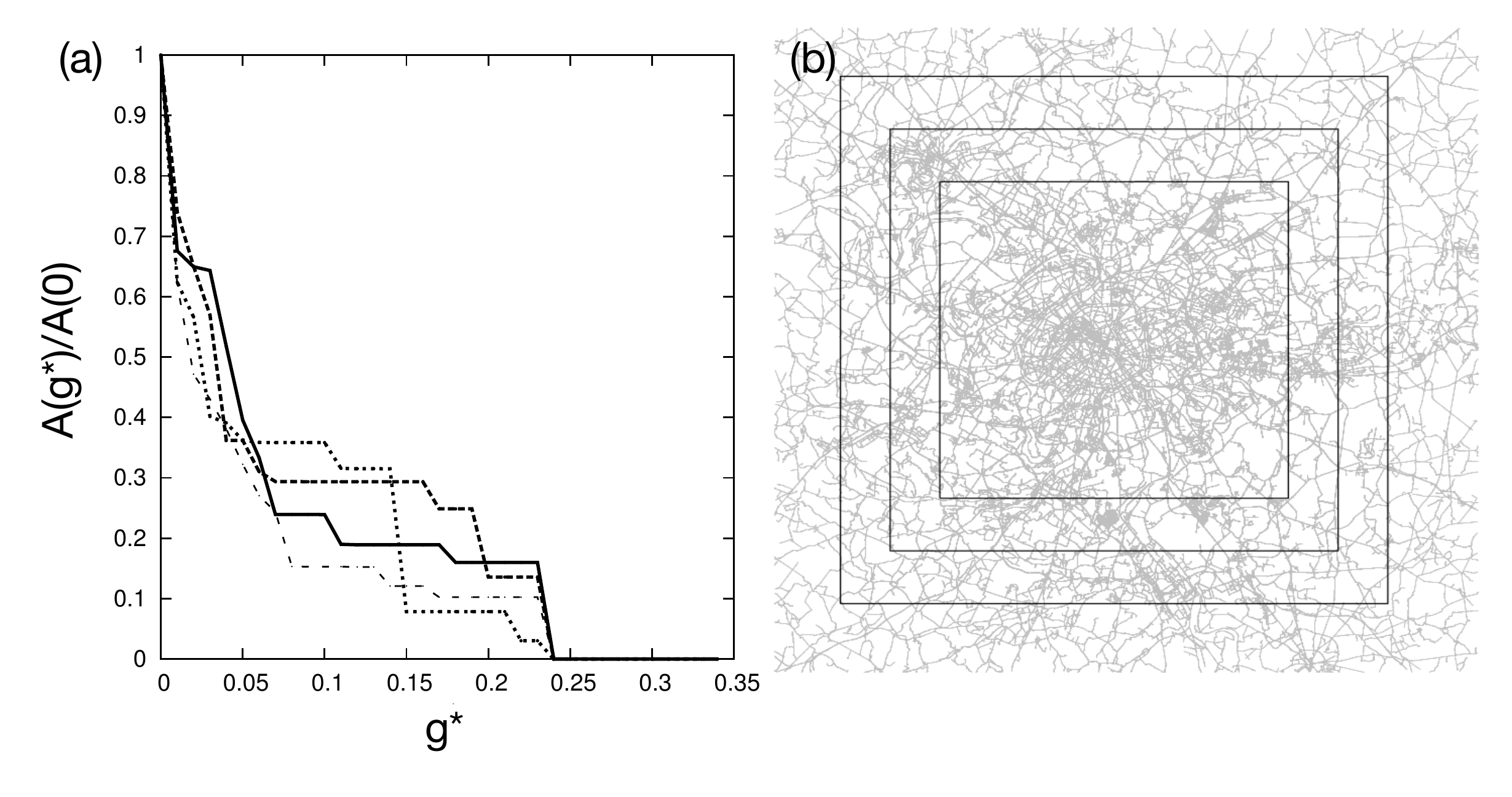}
\caption{ (a) Normalized area $A(g^*)$ defined by the largest loop for
  different boundary conditions on the Paris road network. The lowest
  curve corresponds to the largest size and for decreasing size the
  curves are shifted to larger values of the area. (b) Different
  boundaries corresponding to the curves of (a).
\label{fig:boundary}}
\end{figure}

In general, boundary effects can be important and can affect the
measures done on spatial networks \cite{Rheinwalt:2012}. In
general, the choice of boundaries has an impact on quantities such as
the BC \cite{Gil:2016} and we briefly discuss this problem here. 
We measure the area enclosed in the largest loop on the same network but at different scales
(ie. with different boundaries, going from central Paris to almost the
whole Ile-de-France region) and the results are presented in
Fig. \ref{fig:boundary}. 
In this figure, we observe that at least the area of the largest loop
remains relatively stable. Further systematic studies are however
certainly needed in order to understand which patterns are stable and
which ones are not, and what are the conditions on the boundaries in
order to ensure stability of the main spatial patterns.

\subsection{Summary: stylized facts}

These different examples discussed above show that the introduction of
disorder in planar graphs induce in general the formation of
non-trivial structures made of links with a large BC. In particular, we
observe the appearance of loops made of links that can have a BC
larger than the barycenter. In other words, disorder can invert the
typical behavior observed for regular lattice where the BC is decreasing
monotonously from the barycenter. In the following, we propose a toy
model which allows to discuss and to understand under which conditions
a loop can become more central than the spatial center.

\section{Theoretical approach: A toy model}

\begin{figure}[ht!]
\centering
\includegraphics[width=0.4\textwidth]{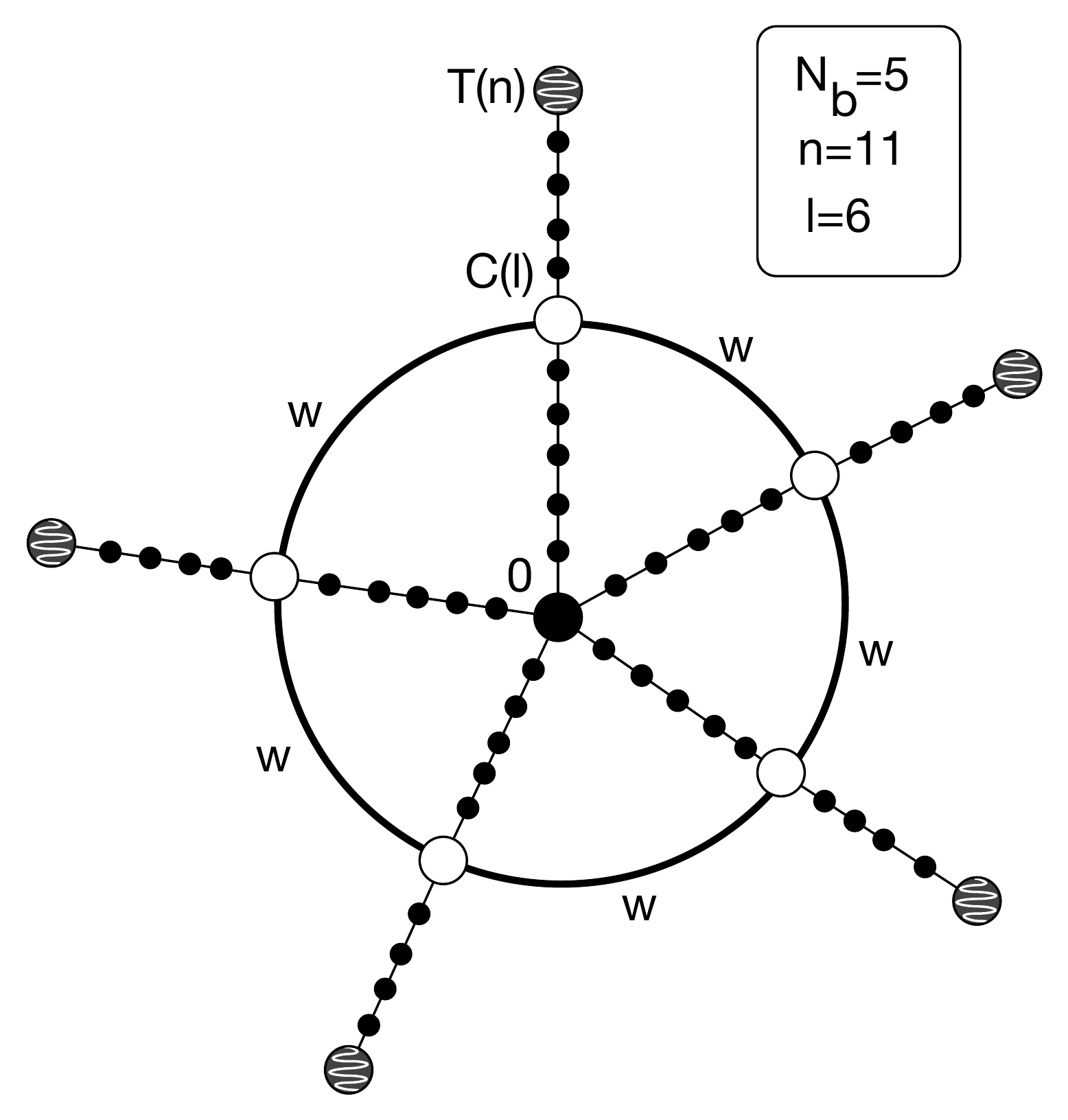}
\caption{ Representation of the toy model discussed
  here. The number of branches is here $N_b=5$, the number of nodes on
  each branch is $n=11$ and the loop is located at a distance $\ell=6$
  from the center $0$. The node $C$ is at the intersection of a branch
and the loop and $T$ is the terminal node of a branch.}
\label{fig:toymodel}
\end{figure}

As discussed above, we observe that non-trivial objects such as loops
can be very central in random graphs. It is important to understand
the formation of these structures and the conditions for their
existence.  In particular, it seems that randomness can induce very
large perturbation in the spatial distribution of the BC and where the
barycenter is not the most central node. Equivalently, the BC could
not be a simple decreasing function of the distance to the barycenter
anymore. In order to understand this phenomenon, we propose here a
simple toy model. We first construct a star network composed of $N_b$
branches, where each branch is composed of $n$ nodes. We then add a
loop at distance $\ell$ from the center (see Fig.~\ref{fig:toymodel}
for a sketch of this graph). We also consider here a more general case
where the links are weighted and in this simplified model we assume
that links have a weight equal to one and the loop segment between two
consecutive branches has a weight given by $w$. The purely topological
case then corresponds to the case $w=1$. We then compute the BC using
weighted shortest paths. This generalization allows us to discuss for
example the impact of different velocities on a street network. In
this case, $w$ can be seen as the time spent on the segment and the
weighted shortest path is then the quickest path.

Here, we want to discuss under which conditions the loop will be more
central than the `origin' at the center in this simplified
network. Intuitively, for very large $w$, it is always less costly to
avoid the loop, while for $w\to 0$, loops are very advantageous.  The
two main quantities of interest are therefore the centrality at the
center denoted by $g_0(\ell,n,w)$ and the centrality, denoted by
$g_C(\ell,n,w)$, at the intersection $C$ of the branch and the
loop. We then compute the difference $\delta g=g_0-g_C$ and will study
under which condition it can be negative.

\subsection{Exact and approximated formulas}

The interest of this toy model lies in the fact that we can estimate
analytically the BC for the center $g_0(\ell,w)$ and for the
intersection nodes on the loop $g_C(\ell,w)$. Formally we can write
these quantities as
\begin{align}
\nonumber
g_0(\ell,n,w)&=g_0(\ell,n,\infty)-(a_1^0+a_2^0+a_3^0)\\
g_C(\ell,n,w)&=g_C(\ell,n,\infty)+(a_1^C+a_2^C+a_3^C)
\label{eq:exact}
\end{align}
where the $a_x^i$ are positive. We distinguish two parts in these
centralities. First, we estimate the BC when there is no loop which is
represented by the case where $w\to\infty$. This part is modified by
the presence of the loop that under certain conditions can be more
interesting for connecting pairs of nodes. We can understand the signs
in Eq.~\ref{eq:exact}, by noting that the presence of the loop will
decrease the centrality at the center and increase the centrality at
$C$. The different terms $a_i^x$ (where $x=0,C$ and $i=1,2,3$) count
the paths (that avoid $0$) connecting two nodes that lie on different
parts of their branch. We divide the nodes on a branch in two parts -
the lower part comprises all nodes that are `below' the loop
$0<s<\ell$ and the upper part is the rest $\ell<s\leq n$. When both
nodes are on the upper part of the branches we obtain $a_1^x$; the
paths connecting an upper part to a lower part are described by
$a_2^x$ and when both nodes lie on a lower part, we obtain the
coefficient $a_3^x$. For more details and the calculation of these
coefficients, we refer to the appendix.

The exact expressions for the centralities $g_0$ and $g_C$ are however
difficult to handle analytically, essentially because they are
expressed as sums of complicated arguments (see appendix). In order to
derive analytical predictions we will propose in the following a
simple approximation scheme that allows to obtain the correct scalings
for the most important quantities. 

\begin{figure}[!h]
\centering
\includegraphics[scale=0.4]{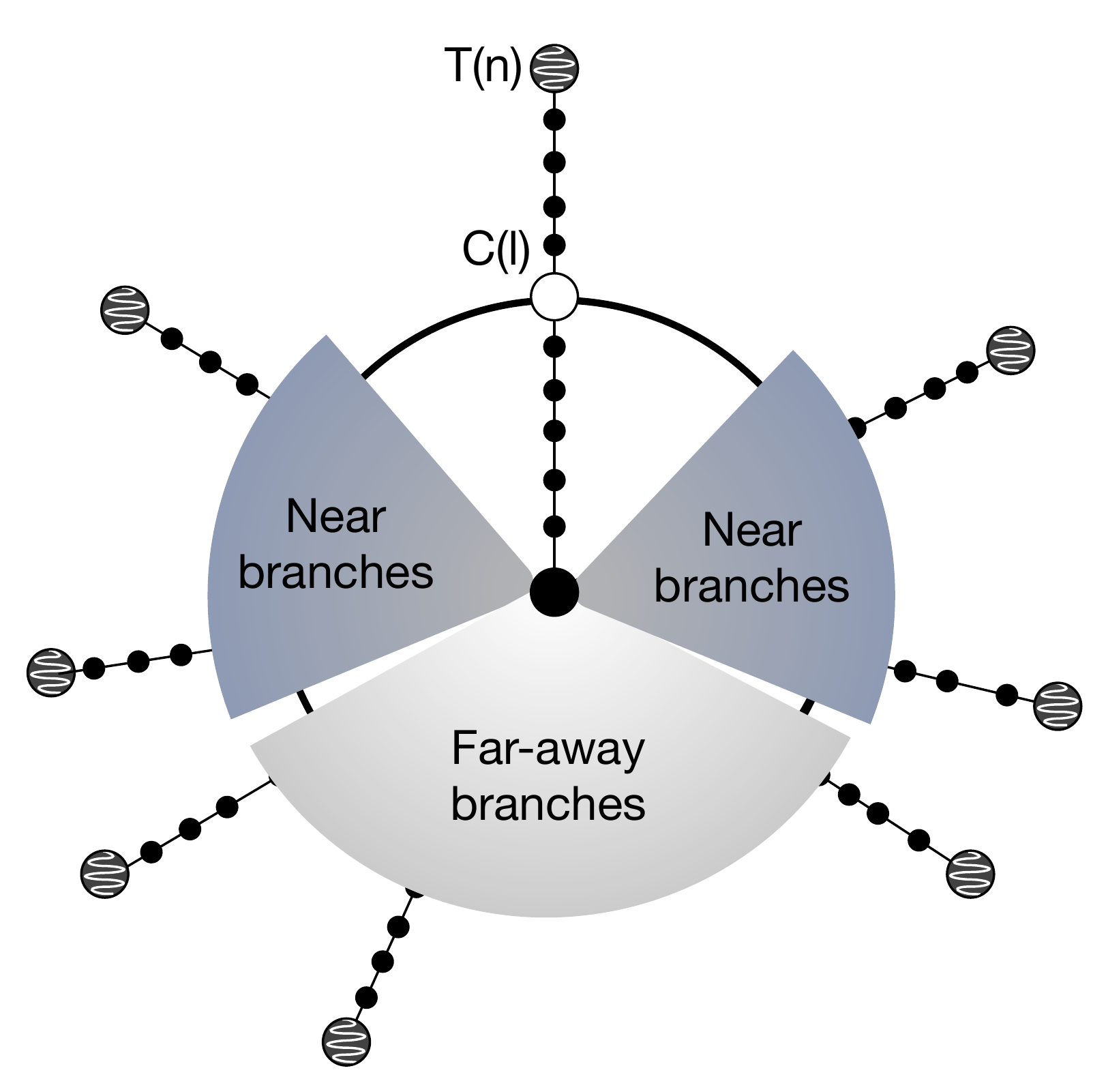}
\caption{ Schematic representation of the approximation used to
  compute the centrality $g_0(w)$ at the center $0$. \label{fig:approx_g0}}
\end{figure}

In the derivation of the exact expression of the centralities
Eq.~\ref{eq:exact}, we have to distinguish different cases according
to the value of
\begin{equation}
\chi\equiv
\min\left(\frac{N_b-1}{2},\left[\frac{2\ell}{w}\right]\right)
\label{eq:chi}
\end{equation}
compared to $j-1$ (the brackets $[\cdot]$ denote here the integer
part, ie. the lowest nearest integer) which denotes the number of loop
segments between the first branch and the branch $j$. This essentially amounts to compare the cost of the
path between a node on the lower part (with $0<s<\ell$) of the first branch $B_1$ to a node on
the lower part ($0<t<\ell$) of another branch $B_j$. If $[2\ell/w]>j-1$ the cost of
the path which goes through $0$ is larger than going directly via the
loop (given by $(j-1)w$) and therefore produces a negative contribution to $g_0$. We see
that this discussion allows to distinguish \- for a given value of $w$
\- `near' from `far-away' branches (Fig.~\ref{fig:approx_g0}). The
nearest branches are then defined by the condition $j-1\leq\chi$ and the
remote branches by $\chi<j-1\leq (N_b-1)/2$ (for
simplicity we assume here that $N_b$ is odd and by symmetry we can
discuss only one half for the branches from $j=2$ to $j=(N_b-1)/2$). 
We will then use the following simplification: we will assume that for
the $\chi$ near branches, going through the center is always the best
choice for $s<\ell$ and $t<\ell$ while for all other paths (from or to
an upper path) it is beneficial to go through the loop. This leads
then to
\begin{equation}
g_{near}=\chi\frac{(\ell-1)(\ell-2)}{2}
\end{equation}
For the $(N_b-1)/2-\chi$ far-away branches, we consider that for all nodes
$s,t\in \llbracket 1,n \rrbracket$, the paths are going through the center leading to
\begin{equation}
g_{far}=\left(\frac{N_b-1}{2}-\chi\right)n^2
\end{equation}
Taking into account the factor $2$ for not counting twice the same
path, we obtain for $g_0(w)=N_b(g_{near}+g_{far})$ the following expression 
\begin{equation}
g_{0}(w)\approx N_b\left\{\left(\frac{N_b-1}{2}-\chi\right)n^2+\chi\frac{(\ell-1)(\ell-2)}{2}\right\}
\end{equation}
We note that this approximation recovers both exact limits
\begin{equation}
g_{0}\simeq
\begin{cases}
N_b\frac{(N_b-1)}{2}\frac{(\ell-1)(\ell-2)}{2} & \;\mathrm{for}\;w\to 0\\
n^2N_b\frac{(N_b-1)}{2} & \;\mathrm{for}\;w\to\infty
\end{cases}
\end{equation}
In the following it will also be useful to consider the limit
$\ell,n\to\infty$ with $x=\ell/n$ fixed which gives for
$g_0(x,\chi)=g_0(\ell,n,w)/n^2$ (up to terms of
order $1/n$)
\begin{equation}
g_{0}(x,\chi)\approx N_b\left\{\left(\frac{N_b-1}{2}-\chi\right)+\frac{1}{2}\chi
  x^2\right\}
\label{eq:g0approx}
\end{equation}
where the only dependence on $w$ is now encoded in $\chi$,
hence the change of argument for clarity.

We can produce the same type of arguments for the BC on the
loop. First the value without the loop is easy to compute and we obtain
\begin{equation}
g_C(\ell,n,w=\infty)=(n-\ell)\left[\ell+n(N_b-1)\right]
\end{equation}
which simply counts the number $(n-\ell)$ of nodes `above' $C$ and all 
the others ($C$ being excluded). Similar arguments as above then give
the following result (we also changed here the argument of the
function from $w$ to $\chi$)
\begin{align}
\nonumber
g_C(\ell,n,\chi)&=g_C(\ell,n,w=\infty)\\
\nonumber
&+2\chi\Big[(n-\ell+1)(\ell-1)+\frac{(\ell-1)(\ell-2)}{2}\Big]\\
\nonumber
&+\frac{\chi(\chi-1)}{2}\Big[(n-\ell+1)^2\\
&+2(n-\ell+1)(\ell-1)+\frac{(\ell-1)(\ell-2)}{2}\Big]
\end{align}
where $\chi$ is given by Eq.~\ref{eq:chi}. In particular the term proportional to $\chi$ counts all the paths
between the lower part of the branch containing $C$ and all the nodes of a branch close
enough. The second term (proportional to $\chi(\chi-1)$) counts the
paths going from a branch $B_j$ with $j\in[1,\chi-1]$ to the other
branches $j'=1,2,\dots,j-1$. The sum of all these contributions gives the
factor $\chi(\chi-1)/2$. The counting factor is not trivial here and
comes from evaluating all the paths from a node $s$ in a branch $j$ to
a node $t$ on a branch $j'$ ($j$ and $j'$ are different from $1$) such 
that
\begin{equation}
s+t>|\ell-s|+|\ell-t|+w\Delta j
\end{equation}
The left hand side of this inequality corresponds to the distance
through the center and $w\Delta j$ is the distance on the loop (for
the exact expression of the centrality and how to recover this
approximate formula, we refer the interested reader to the appendix).

Similarly to the case of the BC at $0$, it will be convenient for analyzing these
expressions to consider the limit $n,\ell\to\infty$ such that
$\ell/n=x$. Up to terms of order $1/n$ we then obtain for
$g_C(x,\chi)=g_C(\ell,n,w)/n^2$
\begin{align}
\nonumber
g_C(x,\chi)&=(1-x)(x+N_b-1)\\
\nonumber
&+2\chi  x(1-\frac{x}{2})\\
&+\frac{\chi(\chi-1)}{2}(1-\frac{x^2}{2})
\label{eq:gCapprox}
\end{align}

\begin{figure}[!h]
\centering
\includegraphics[scale=0.5]{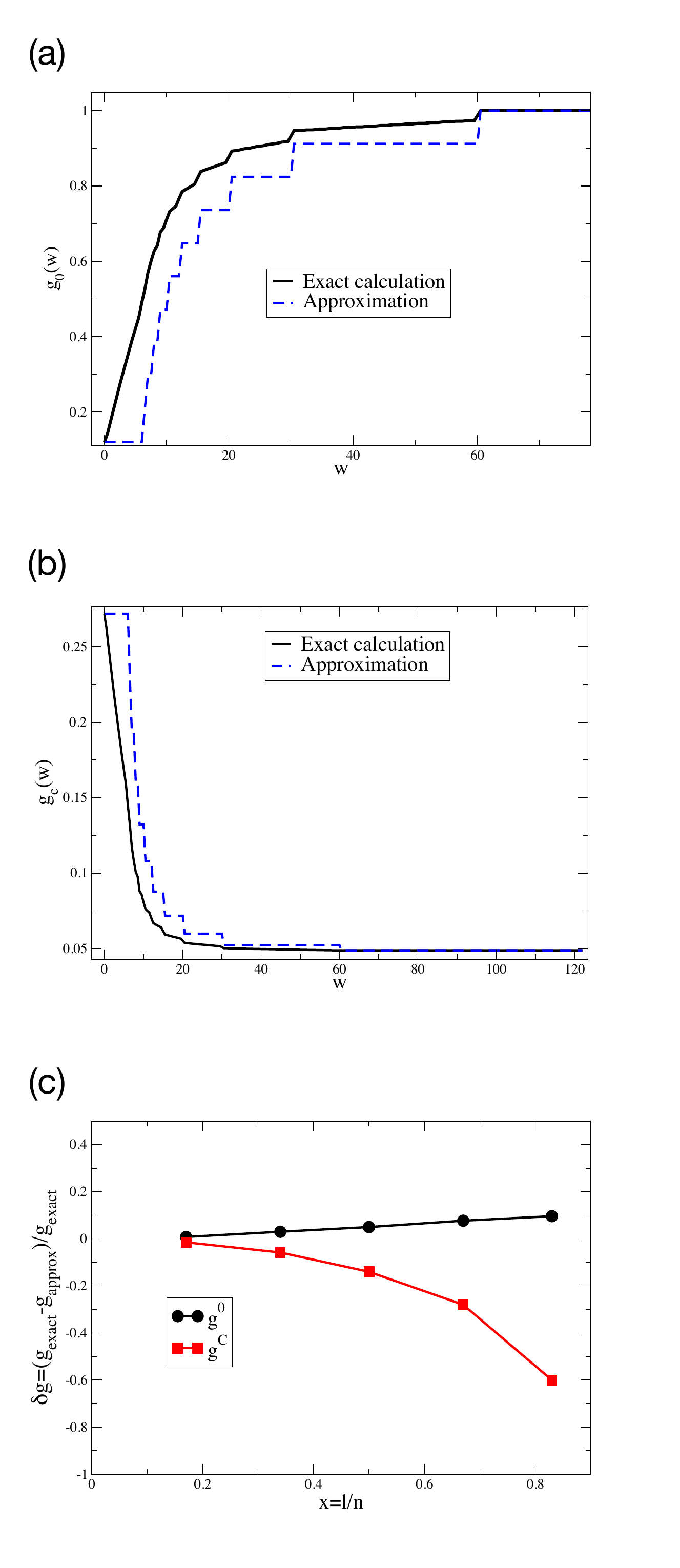}
\caption{ Comparison between the exact result and the approximation
  for $g_0(w)$ (a)  and $g_C(w)$ (b) (the BC are here
  normalized). The parameter values are
  here $N_b=21$, $n=60$ and $\ell=30$. (c) Relative error between
the exact value and approximation for $g_0$ and $g_C$ for $N_b=21$,
$n=60$, $w = 30$, $\ell=30$.}
\label{fig:approx}
\end{figure}

We show in the figure \ref{fig:approx} the comparison of the exact
result with the approximations developed here. 
For large values of $\ell$ the approximation is not excellent and
can certainly be improved. However as we will show in the following,
our simple approximations allow to understand and to predict the
correct scaling for the important quantities $\ell_{opt}$ and $w_c$.

\subsection{Threshold value of $w$ and optimal $\ell$}

\begin{figure}[!h]
\centering
\includegraphics[scale=0.30]{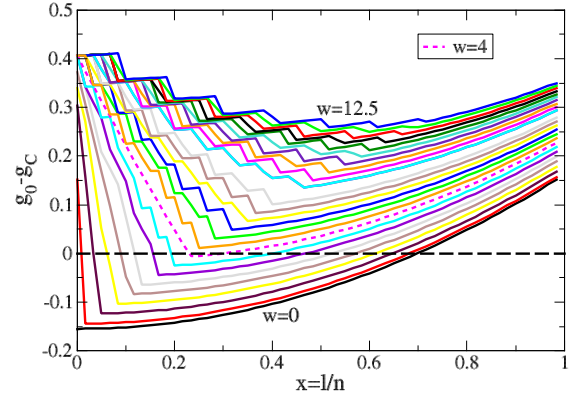}
\caption{ $\delta g(\ell)$ versus $\ell$ (for $N_b=15$ and $n=60$ here) and
  for different values of $w$ in the range $[0,12.5]$. For values less
  than a threshold ($w_c\approx 4$ shown here by a dotted line)  there is a minimum that is negative.
\label{fig:deltag}}
\end{figure}

The fundamental quantity that we wish to understand is the difference
$\delta g(x,\chi)=g_0(x,\chi)-g_C(x,\chi)$ given by Eqs.~(\ref{eq:g0approx},\ref{eq:gCapprox}).
We first plot this quantity versus
$\ell$ for different values of $w$ and we observe the result shown in
Fig.~\ref{fig:deltag}
This result shows that for $w$ sufficiently small, $\delta g$ can be
negative. This demonstrates the existence of a threshold value $w_c$
such that at $w=w_c$ the minimum is $\min_\ell\delta g=0$. 
For $w<w_c$, the minimum of $\delta g$ is negative and we can define
an optimal value $\ell_{opt}$ which corresponds to this smallest value
of $\delta g$. The quantity $\ell_{opt}$ thus gives the position of
loop that maximizes the difference between the BC of the loop and
the center. 

In order to estimate this optimal value $\ell_{opt}$, we note (using
the expression Eq.~\ref{eq:chi} for $\chi$) that the difference
$\delta g(x,\chi)$ gives
\begin{equation}
\delta g(x,\chi)=
\begin{cases} 
\delta g(x,\frac{2\ell}{w}) & \;\mathrm{for}\;\ell\in[0,\frac{(N_b-1)w}{4}]\\
\delta g(x,\frac{N_b-1}{2}) & \;\mathrm{for}\;\ell\in[\frac{(N_b-1)w}{4},2n]
\end{cases}
\end{equation}
In order to discuss to estimate analytically both the threshold $w_c$
and the optimal value $\ell_{opt}$, we will use equations
Eqs. (\ref{eq:g0approx},\ref{eq:gCapprox}) and study the approximate
difference $\delta g(x,\chi)=g_0(x,\chi)-g_C(x,\chi)$ given by
\begin{align}
\nonumber
\delta g(x,\chi)&=N_b\left[\frac{N_b-1}{2}-\chi+\frac{1}{2}\chi
  x^2\right]\\
\nonumber
&-(1-x)(x+N_b-1)-2\chi x(1-\frac{x}{2})\\
&-\frac{\chi(\chi-1)}{2}(1-\frac{x^2}{2})
\label{eq:dgall}
\end{align}

\begin{figure}[!h]
\centering
\includegraphics[scale=0.3]{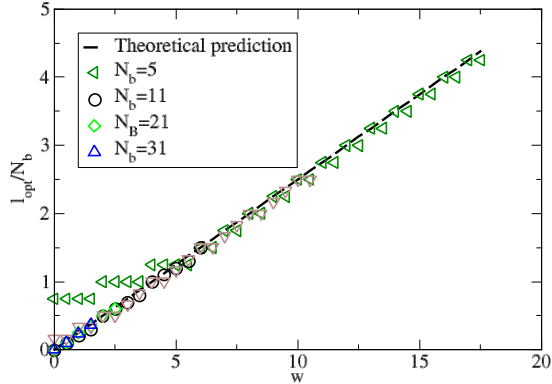}
\caption{ Comparison between the theoretical prediction
  (Eq.~\ref{eq:lopt}) and numerical
  results for $\ell_{opt}$ (for $w<w_c$). For large value of $N_b$ the prediction is excellent. We
  note that $\ell_{opt}$ exists for $w<w_c$ and $w_c$ decreases with
  $N_b$ which implies that the range over which we can see a linear
  behavior is decreasing as $1/N_b$ (here $n=40$). }
\label{fig:lopt}
\end{figure}

We first study the derivative with respect to $\ell$ of this
difference in the domain $\ell <[(N_b-1)w/4]$. After simple
calculations we obtain that for large $N_b$ and $n$ (we treat here
$\ell$ as a continuous variable) $\mathrm{d}\delta g/\mathrm{d}\ell<0$
in the domain considered. A similar calculation shows that in the
domain $(N_b-1)/4<\ell<2n$, the function $\delta g(\ell,n,\chi)$ is
increasing with $\ell$ (at least for $N_b$ large enough). These
results thus show that the minimum of $\delta g$ is actually reached
at the intersection of the two curves and which occurs for
\begin{equation}
\ell_{opt}=\frac{(N_b-1)w}{4}
\label{eq:lopt}
\end{equation}
This expression for $\ell_{opt}$ is actually independent from the exact form of $\delta
g$ as long as it is decreasing for $\ell<\ell_{opt}$ and increasing above
$\ell_{opt}$ \- which we verified numerically. We compare the
theoretical prediction Eq.~(\ref{eq:lopt}) with numerical results in
Fig.~\ref{fig:lopt}, 
and we see that for $N_b$ large enough (here, typically $N_b>10$) this prediction is
in excellent agreement with data. 

We can understand this value of $\ell_{opt}$ with the following simple
argument. If $\ell$ is small most paths connecting nodes from
different branches will go through $0$ and we expect $\delta
g>0$. When $\ell$ is increasing more paths will go through the loop and will
increase the value of $g_C$. However, when $\ell$ is too large, paths
connecting the (large) fraction of nodes located on the lower branches
will go through $0$ again. In order to get a sufficient condition on
$\ell_{opt}$, we consider the path between the node $C$ on the branch
$B_1$ and the corresponding node $C'$ on the furthest branch
$(N_b-1)/2$. The optimal value for $\ell_{opt}$ is then such that the
cost of the path from $C$ to $C'$ through $0$ and which is $2\ell$ is
equal to the cost on the loop which is given by $w(N_b-1)/2$. This
immediately gives the result $\ell_{opt}\approx w(N_b-1)/4$.

The threshold quantity $w_c$ is obtained by imposing that the minimum
of $\delta g(\ell=\ell_{opt})$ is equal to zero. Using the approximate
form Eq.~\ref{eq:dgall}, we can show that the minimum is actually obtained for
$\ell=\ell_{opt}$ and for $\chi=(N_b-1)/2$. We thus have to consider
the quantity $\delta g(\ell_{opt},n,\chi=(N_b-1)/2)$ which for large
$N_b$ is behaving as
large $N_b$
\begin{equation}
\delta g(\ell_{opt})\approx
%\frac{N_b^2}{8}\left[\frac{3}{2}\left(\frac{wN_b}{4n}\right)^2-1\right]
\frac{N_b^2}{8}\left[\frac{5}{2}\left(\frac{wN_b}{4n}\right)^2-1\right]
\end{equation}
(details of this calculation are given in appendix) and we therefore obtain
\begin{equation}
w_c\approx \kappa\frac{n}{N_b}
\label{eq:wc}
\end{equation}
 where $\kappa=4\sqrt{\frac{2}{5}}$ in this approximation.
\begin{figure}[!h]
\centering
\includegraphics[scale=0.3]{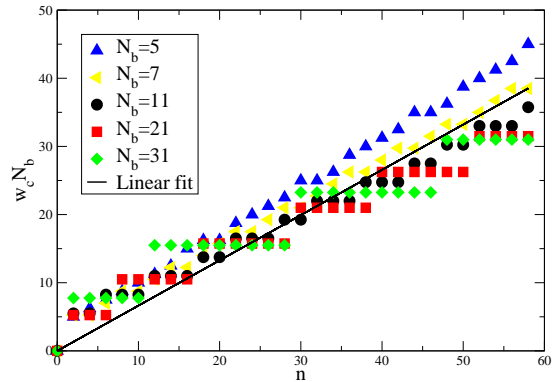}
\caption{ Value of $w_cN_b$ versus $n$. The collapse is reasonably
  good and is in agreement with our theoretical result Eq. \ref{eq:wc}. We observe plateaus
  that are due to the discrete values of $\ell$ and $n$. The straight line is a linear fit which gives
  $\kappa_{emp}\approx 0.66$ ($r^2=0.96$).}
\label{fig:wc}
\end{figure}
We can understand the scaling for $w_c$ with the simple following
argument. Indeed, a necessary condition on $w$ is that $\ell_{opt}$
must be less than $n$. This gives the condition
\begin{equation}
w<\widetilde{w}_c=4\frac{n}{N_b}
\end{equation}
This threshold $\widetilde{w}_c$ is a priori larger than the exact value,
as we imposed here a necessary condition, but allows to understand in
a simple way the scaling of $w_c$ with $n$ and $N_b$. We test the
scaling for $w_c$ by plotting (Fig.~\ref{fig:wc}) $w_cN_b$ versus $n$
and which should be linear. We indeed observe a reasonable agreement
with the linear behavior predicted by our analysis, where the
differences are probably due to the small values of $N_b$ used for the
numerical calculations. The linear fit however gives a prefactor
$\kappa_{emp}\approx 0.66$ which is far from the value obtained within
our simple approximation scheme. The important fact is that our
approximation is able to predict the correct scaling and it could
maybe be possible to find more refined approximations in order to get
a better estimate for the prefactor $\kappa$. Finally, we note here
that $w_c$ is independent from $\ell$ which can be understood by the
fact that $w_c$ gives a condition for the existence of $\ell_{opt}<n$.

Finally, we note that when $w_c>1$, the case $w=1$ displays then a
negative minimum and we can observe a very central loop. This case is
particularly interesting as it corresponds to the `topological' case
for which the distance is the minimum number of hops. This will then
happen when there are few branches, or if the branches are large
enough.

\section{Discussion}

The main purpose of this paper is to shed light on the appearance of
non-trivial patterns made of very central nodes (or links) in
real-world planar graphs. In particular, we focused on the existence
of very central loops that are commonly observed in random planar
graphs. We proposed a toy model that shows that indeed a loop at a
certain distance from the center can be more central than the physical
center itself. The condition for the existence of such a phenomenon is
that the weight on the loop has to be small enough and in this toy
model we showed the existence of a threshold value $w_c$. This
threshold depends on the size and number of radial branches,
highlighting their crucial role. This result allows us to understand
the appearance of very central loop even in the topological case where
the shortest topological distance is used for computing the BC: if the
extension of the network is large compared to the number of radial
branches, $w_c$ can be larger than one $w_c>1$ and central loops for
$w=1$ can be observed. In ordered systems -- such as lattices -- the
effective number of branches is too large leading to a very small
$w_c$ and therefore prohibits the appearance of central loops in the
`topological' case ($w=1$). In real-world planar graphs where
randomness is present, the absence of some links can lead to a small
number of `effective' radial branches which in the framework of the
toy model implies a large value of $w_c$ and therefore a large
probability to observe central loops.

In the case of roads, if we assume that the weight is
inversely proportional to the velocity, our result predicts that the
velocity on the loop has to be large enough in order to be very
central and faster than going through the center. A possible direction
of future studies could then to include more precisely different
velocities and also to include congestion effects. 

More generally, further studies are needed in order to understand the
variety of patterns induced by randomness in planar graphs, and we
believe that this study is a step towards this direction.

%Indeed if we denote by $v$ the velocity on radial road and $v'$ the velocity on
%loops (we assume that $v'>v$ in order to make the loop interesting), the
%value of $w$ corresponds to the ratio of the time $\tau_b$ spent on an
%elementary segment on branches divided by the time $\tau_l$ spent on a loop segment
%between two branches. If the length of branches is given by $R$ we
%then have $\tau_b\sim R/nv$. The loop segment has a radius given by
%$R\ell/n$
%and we thus have $\tau_l\sim \frac{2\pi}{N_b} R\ell/nv'$. The value $w$ corresponds to
%the ratio of these times and we have
%\begin{equation}
%w=\frac{\tau_l}{\tau_b}\sim\frac{2\pi}{N_b}\frac{v}{v'}n\ell
%\end{equation}
 %Writing that $w<w_c$ then gives a condition on the velocity ratio
%\begin{equation}
%\frac{v}{v'}<\frac{\kappa}{2\pi}\frac{1}{\ell}
%\end{equation}
%There is therefore a minimal velocity for the loop to be more central, and
%therefore attractive from a time cost point of view, but the existence
%of the loop is not enough to guarantee its potential. 

\section{Acknowledgements}

BL and MB thank Jean-Marc Luck for interesting discussions and valuable
suggestions. MB thanks Gourab Ghoshal for stimulating exchanges on
this problem.

\section{Appendix}

\subsection{BC at the center}

\subsubsection{Complete formula}

The general expression for the BC at the center is
\begin{equation}
g_0(w)=n^2N_b(N_b-1)/2-(a_1^0+a_2^0+a_3^0)
\end{equation}
In order to analyze this quantity, we separate the branches into two
parts. The first part is composed of nodes at a distance lower than
$l$ from the center (the `lower' part of the branch), and the second
part consists in nodes that are at a distance greater or equal than
$l$ from the center (the `upper' part of the branch). When there is no
loop, the total number of shortest paths going through the center is
($N_{tot}$ denotes the total number of nodes)
\begin{align}
\nonumber
g_{0}(w=\infty)&=\dbinom{N_{tot}-1}{2}-\dbinom{n}{2}\cdot{N_{b}}\\
&=n^2N_b(N_b-1)/2
\end{align}
This expression gives the betweenness centrality at the center without
the loop, and we can calculate $g_0(w)$ by removing from $g_0(\infty)$
all the shortest paths that go through the loop and not through the
center. This can be computed by distinguishing different types of
paths: the quantity $a_1^0$ counts the number of shortest paths going
through the loop and connecting nodes both located on the upper part
of branches; the quantity $a_2^0$ counts these paths connecting an
upper part to a lower one; and $a_3^0$ counts the paths connecting
nodes both located in lower parts of branches.

We note that due to the symmetry of this network, it is enough to
consider paths from a given branch to the others and to multiply at
the end of the calculation by the number of branches.

The central point for calculating the centralities $g_0(w)$ and
$g_C(w)$ is to compare the length of
the path through $0$ and the path through the loop. For a node located
at distance $s$ on branch $0$ and a node at distance $t$ on a branch
$j$ (where $j$ goes from $1$ to $N_b-1$), this condition can be written as
\begin{equation}
s+t<|\ell-s|+|\ell-t|+wj
\end{equation}
If this inequality is satisfied the path will go through $0$ and
otherwise the loop is more interesting. We thus have to count the pair
of nodes that satisfies this inequality and for this we distinguish
three different cases: $s$ and $t$ in the lower part of branches, $s$
and $t$ in the lower and upper parts respectively, and finally both
$s$ and $t$ in the upper part of branches. In the following we
introduce two quantities:
\begin{equation}
X_j=\min\left(\left[\frac{2j}{w}\right],\frac{N_b-1}{2}\right)
\end{equation}
and 
\begin{align}
\nonumber
P_j&=\frac{1}{2}\theta(N_b\;\;\mathrm{odd})\theta(\frac{2j}{w}=X_j)\\
\nonumber
&+\theta(N_b\;\;\mathrm{even})\left(\frac{2}{3}\theta(\frac{N_b}{2}=\frac{2j}{w})+\frac{1}{2}\theta(\frac{N_b}{2}<\frac{2j}{w})\right)
\end{align}
where
$\theta(condition)=1$ if condition is true, and is $0$ otherwise. This
quantity $P_j$ takes into account the fact that the shortest path
going through the center or through the loop have the same length. If $2j/w=X_j$ we
have to divide by two the number of paths and $X_j-P_j$ is the
fraction of paths going through the center. In addition, if $N_b$ is
even and $N_b/2=2j/w$, we have $3$ different paths with the same length that connect two
nodes on opposite branches: one path through the center, another 2
paths on each direction of the loop. In this case, there are $2/3$ of
paths to remove in order to get $g_0$. in each case $X_j-P_j$
multiplies the total flow.

\medskip
\paragraph{Calculation of $a_1^0$}

There are $(n-\ell+1)$ nodes in the upper part of one
branch and therefore $N_b(n-\ell+1)^2$ possible pairs between two nodes
of two distincts upper parts. The number of branches with nodes that
will deviate from the center and will use the loop is given by
$X_\ell$. We have however to consider cases where there are shortest
paths that equivalently go through the center or via the loop, and
this is precisely what is counted by $P_\ell$. The coefficient $a_1^0$
is then given by
\begin{equation}
a_1^0=N_b(n-\ell+1)^2(X_\ell-P_\ell)
\end{equation}
We can recover this result by noting that for nodes $s$ and $t$
belonging to the upper part of different branches $B_1$ and $B_j$, the
condition that the path through the center is longer than on the loop
is
\begin{equation}
s+t>s-\ell+t-\ell+wj
\end{equation}
which leads to $2\ell>wj$ and a number of branches given by 
$[2\ell/w]=X_\ell$. We have an equality when $2\ell=wj$ and this
happen when $X_\ell=2\ell/w$ and gives a factor $1/2$ in the BC.

\medskip
\paragraph{Calculation of $a_2^0$}

For the $a_2^0$ coefficient, only paths between the lower part and the
upper part are considered. We consider a node (in the first branch) in
the lower part $s<\ell$ and another one in the upper part on $t>\ell$
in the branch $j$. The path from $s$ to $t$ is deviated from zero if
the following condition is met
\begin{align}
\nonumber
&s+t>|\ell-s|+|\ell-t|+wj\\
&\Rightarrow 2s>wj
\end{align}
which means that the number of such paths is given by the number of nodes
in the upper part $(n-\ell+1)$ times the number of branches that
satisfy this condition: $j<\left[\frac{2s}{w}\right]$. We have to sum
over $s\in\llbracket 1,\ell\rrbracket$ and to 
multiply by $2N_b$ which takes into account both paths (from the upper to
the lower and from the lower to the upper part of branches) and obtain
\begin{equation}
a_2^0=2N_b(n-\ell+1)\sum_{s=1}^{\ell-1}(X_{s}-P_{s})
\end{equation}
(the term $P_s$ takes into account the degeneracies of paths).

\medskip
\paragraph{Calculation of $a_3^0$}

The quantity $a_3^0$ represents deviation between pairs of nodes both
located in the lower part of branches. in this case, the condition on
paths is
\begin{equation}
s+t>\ell-s+\ell-t+wj
\end{equation}
which implies that the branches where we have a deviation from $0$ are
such that
\begin{equation}
j<\frac{2(s+t-\ell)}{w}
\end{equation}
and we then obtain
\begin{equation}
a_3^0=N_b\sum_{s=1}^{\ell-1}\sum_{t=1}^{s-1}(X_{s+t-\ell}-P_{s+t-\ell})
\end{equation}

We therefore finally obtain the exact expression
\begin{equation}
g_0(w)\approx n^2N_b(N_b-1)/2-(a_1+a_2+a_3)
\end{equation}
with
\begin{align}
&a_1=N_b(n-\ell+1)^2 \cdot (X_\ell-P_\ell)\\
&a_2=2N_b(n-\ell+1)\sum_{s=1}^{\ell-1}(X_{s}-P_s)\\
&a_3=N_b\sum_{s=1}^{\ell-1}\sum_{t=1}^{s-1}(X_{s+t-\ell}-P_{s+t-\ell})
\end{align}
where $X_j=\min(\left[\frac{2j}{w}\right],\frac{N_b-1}{2})$ and $P_j$
given above. The sums entering these expressions are however difficult
to handle and we therefore resort to approximations that are detailed
in the main text.

\medskip
\subsubsection{Simplification}
\medskip

Another way to recover the approximation discussed in the main text is
to neglect small deviation terms (ie. to impose $P_j= 0$), and we then obtain
\begin{equation}
g_0(w)\approx n^2N_b(N_b-1)/2-(a_1+a_2+a_3)
\end{equation}
with
\begin{align}
&a_1=N_b(n-\ell+1)^2 \cdot X_\ell\\
&a_2=2N_b(n-\ell+1)\sum_{s=1}^{\ell-1}X_{s}\\
&a_3=N_b\sum_{s=1}^{\ell-1}\sum_{t=1}^{s-1}X_{s+t-\ell}
\end{align}
We are still left with the sums to compute and the simplest approximation
we can think of (and that can be used for $g_C$ too) is to choose $X_j\approx X_\ell=\chi$ leading to
\begin{align}
&a_1\approx N_b(n-\ell+1)^2 \chi\\
&a_2\approx 2N_b(n-\ell+1)\chi (\ell-1)\\
&a_3\approx N_b\chi \frac{(\ell-2)(\ell-1)}{2}
\end{align}
leading to the result Eq. \ref{eq:g0approx} (in the limit
$n,\ell\to\infty$ with $x=\ell/n$ fixed). This is obviously a very
crude approximation and it could certainly be refined in order to get a
more accurate expression for $g_0$. However, as we will see in the
next section, the expression for $g_C$ is much more involved and we
need an approximation scheme that can be applied to both quantities
$g_0$ and $g_C$, which seems to be a difficult task that we leave for
future studies.

\subsection{BC for the loop}

\subsubsection{Complete formula}

The first term $g_C(w=\infty)=(n-\ell)(n(N_b-1)+\ell)$ is the number of shortest path
passing through $C$ when there is no loop. The quantity $(n-\ell)$ is the
number of nodes in the upper part of the branch $B_1$ and
the number of node in the rest of the network is $(n(N_b-1)+\ell)$. 
Similarly as for $g_0$, we start from the quantity computed for
$w=\infty$ and add the number of paths that will go through the loop
and obtain
\begin{equation}
g_C(w)=(n-\ell)(n(N_b-1)+\ell)+(a_1^C+a_2^C+a_3^C)
\end{equation}

The first term is considering only deviation from the upper part of
the branches to other upper parts. For all other branches
$k=2,\dots,N_b/2$ (by symmetry and for $N_b$ even), if $wk<2\ell$, there are
$((k-1)*(n-\ell+1)*(n-\ell+1))$ additional shortest paths going via the
loop. By summing over $k$ and taking into account
multiple paths, we obtain
\begin{equation}
a_1^C=(n-\ell+1)^2(\frac{1}{2}(X_{\ell-w/2}+1)X_{\ell-w/2}-P_\ell^1)
\end{equation}

The coefficient $a_2^C$ counts the paths from the upper part
of a branch to the lower part of another branch. This path will go
through the loop if $wi<2j$ with $i$ going from $1$ to $N_b/2$ and $j$
from $1$ to $\ell-1$. We then obtain
\begin{equation}
a_2^C=2(n-\ell+1)\sum_{j=1}^{\ell-1}\left(\frac{X_j(X_j+1)}{2}-P_j^2\right)
\end{equation}

Finally, the coefficient $a_3^C$ corresponds to additional shortest paths from lower
part to lower part and going through the loop. When $wi<2(j+k-\ell)$,
with $i=1,\dots,N_b/2$, $j=1,\dots,\ell-1$, and $k$ running from $\ell-j+1$ to $\ell-1$, there are $(i+1)$ new
shortest paths added at point C. Summing over $i$, we then 
obtain 
\begin{equation}
a_3^C=\sum_{j=1}^{\ell-1}\sum_{k=1}^{j-1}\left(\frac{X_{j+k-\ell}(X_{j+k-\ell}+3)}{2}-P_{j+k-\ell}^3\right)
\end{equation}

The quantities $P_j^i,\;i=1,...,3$ correspond to the correction needed when the
path going through the loop has the same weight as the path going
through $0$.

For the part $1$:
\begin{align}
\nonumber
P^1_j&=\frac{X_j}{2}\theta(\frac{2j}{w}=X_j)\theta(\frac{N_b}{2}\neq  X_j)\\
\nonumber
&+\frac{N_b/2-1}{3}\theta(\frac{N_b/2}{2}=\frac{2j}{w})\theta(\frac{N_b}{2}=\left[\frac{N_b}{2}\right]\\
&+\frac{N_b/2-1}{2}\theta(\frac{N_b}{2}<\frac{2j}{w})\theta(\frac{N_b}{2}=\left[\frac{N_b}{2}\right])
\end{align}

For the part $2$:
\begin{align}
\nonumber
P^2_j&=\frac{X_j}{2}\theta(\frac{2j}{w}=x_j)\theta(\frac{N_b}{2}\neq  X_j)\\
\nonumber
&+\frac{2(N_b/2)}{3}\theta(\frac{nb}{2}=\frac{2j}{w})\theta( \frac{N_b}{2}=\left[\frac{N_b}{2}\right])\\
&+\frac{N_b/2}{2}\theta(\frac{N_b}{2}<\frac{2j}{w})\theta(\frac{N_b}{2}=\left[\frac{N_b}{2}\right])
\end{align}

For the part $3$:
\begin{align}
\nonumber
P^3_j&=\frac{X_j+1}{2}\theta(\frac{2j}{w}=X_j)\theta(\frac{N_b}{2} \neq      X_j)\\
\nonumber
&+\frac{2X_j+2}{3}\theta(\frac{N_b}{2}=\frac{2j}{w})\theta( \frac{N_b}{2}=\left[\frac{N_b}{2}\right])\\
&+\frac{X_j}{2}\theta(\frac{N_b}{2}<\frac{2j}{w})\theta(\frac{N_b}{2}=\left[\frac{N_b}{2}\right])
\end{align}

\medskip
\subsubsection{Simplification}

Here also, we can recover the approximate formula
(Eq. \ref{eq:gCapprox}) by using the same approximation as described
above for $g_0$: we neglect small deviation terms ($P_j= 0$) and we
assume that $X_j=X_\ell\approx\chi$ for all
$j$ which gives
\begin{equation}
g_C(w)=(n-\ell)(n(N_b-1)+\ell)+(a_1^C+a_2^C+a_3^C)
\end{equation}
with
\begin{align}
\nonumber
a_1^C&=(n-\ell+1)^2\frac{1}{2}(X_{\ell-w/2}+1)X_{\ell-w/2}\\
&\approx (n-\ell+1)^2\frac{\chi(\chi-1)}{2}\\
\nonumber
a_2^C&=2(n-\ell+1)\sum_{j=1}^{\ell-1}\frac{X_j(X_j+1)}{2}\\
&\approx 2(n-\ell+1)(\ell-1)\frac{\chi(\chi+1)}{2}\\
\nonumber
a_3^C&=\sum_{j=1}^{\ell-1}\sum_{k=1}^{j-1}\frac{X_{j+k-\ell}(X_{j+k-\ell}+3)}{2}\\
&\approx\frac{\chi(\chi+3)}{2}\;\frac{(\ell-2)(\ell-1)}{2}
\end{align}
where $\chi=\min(\left[\frac{2j}{w}\right],\frac{N_b-1}{2})$. Summing
all these terms and taking the limit $n,\ell\to\infty$ with $\ell/n=x$
fixed, we recover the approximation Eq.\ref{eq:gCapprox}.

\subsection{Calculation of $w_c$}

We start with the expression Eq.~\ref{eq:dgall}
\begin{align}
\nonumber
\delta g(x,w)&=N_b\left[\frac{N_b-1}{2}-\chi+\frac{1}{2}\chi
  x^2\right]\\
\nonumber
&-(1-x)(x+N_b-1)-2\chi x(1-\frac{x}{2})\\
&-\frac{\chi(\chi-1)}{2}(1-\frac{x^2}{2})
\end{align}
where $\chi=\frac{N_b-1}{2}$. In the limit $N_b\gg (x-1)$ we obtain
(keeping terms growing with $N_b$)
\begin{align}
\nonumber
\delta g(x,w)\simeq &N_b\left[\frac{N_b-1}{4}x^2\right]\\
\nonumber
&-(1-x)N_b-2\frac{N_b-1}{2}x(1-\frac{x}{2})\\
\nonumber
&-\frac{N_b-1}{4}(\frac{N_b-1}{2}-1)(1-\frac{x^2}{2})
\end{align}
which behaves at leading order as
\begin{equation}
\delta g(x,w)\simeq \frac{N_b^2}{8}\left(\frac{5}{2}x^2-1\right)
\end{equation}

%\nonumber
%\simeq &x^2(\frac{5N_b^2}{16}-\frac{5}{16})\\
%\nonumber
%&+x\\
%\nonumber
%&-\frac{1}{8}(N_b^2-4N_b+3)
%\end{align}
%In the limit $N_b>>1$ and $N_b^2>>N_b$, previous equation gives: 

The minimum is crossing zero at 
\begin{align}
x^2=\frac{2}{5}
\end{align}
which implies for $\ell=\ell_{opt}\simeq N_bw/4$
\begin{equation}
w_c \simeq \sqrt{\frac{2}{5}}\frac{4n}{N_b}
\end{equation}

%%%%%%%%%%%%%%%%% REFERENCES

\bibliographystyle{prsty}

\end{document}